\documentstyle[eqsecnum,aps,amstex,epsfig,float]{revtex}

\begin{document}                

\newcommand{\erfc}{\operatorname{erfc}}

\draft
\title{Detection in coincidence of gravitational wave bursts with a network of interferometric detectors (I): Geometric acceptance and timing}

\author{Nicolas Arnaud, Matteo Barsuglia, Marie-Anne Bizouard, Philippe Canitrot,\\ Fabien Cavalier, Michel Davier, Patrice Hello and Thierry Pradier}

\address{Laboratoire de l'Acc\'el\'erateur Lin\'eaire, B.P. 34, B\^atiment 200,
Campus d'Orsay, 91898 Orsay Cedex (France)\protect\\}

\maketitle

\begin{abstract}
Detecting gravitational wave bursts (characterised by short durations and poorly modelled waveforms) requires to have coincidences between several interferometric detectors in order to reject non stationary noise events. As the wave amplitude seen in a detector depends on its location with respect to the source direction and as the signal to noise ratio of these bursts are expected to be low, coincidences between antennas may not be so likely. This paper investigates this question from a statistical point of view by using a simple model of a network of detectors; it also estimates the timing precision of a detection in an interferometer which is an important issue for the reconstruction of the source location, based on time delays.
\end{abstract}

\pacs{PACS numbers 04.80.Nn, 07.05.Kf}


\baselineskip = 2\baselineskip

\section{ Introduction}
In the next few years, the first generation of long baseline interferometric detectors of gravitational waves (GW) \cite{ligo,virgo,geo,tama} will be operational. Among the most promising sources of GW, the compact binary inspirals and 
the periodic sources have been already studied for a long time (see \cite{thorne87} or \cite{schutz} for a review); more recently, some analysis methods have been developed to look for GW bursts of short duration and unknown waveform which are the subject of this paper.

Type II supernovae (\cite{bona} and references therein) and the merging phase of binary compact star and/or black hole systems \cite{ooh} are the most common burst sources foreseen but others such as e.g. cosmic strings \cite{dv} could also be considered promising. The lack of knowledge of such GW signals prevents from using the optimal (Wiener) filtering method and requires sub-optimal (i.e. less efficient yet robust) filters; various techniques have already been studying by different groups. The ``excess power statistics'' \cite{flana} monitors the power of the detector output along the time and is shown to be optimal when minimal hypothesis (signal duration and bandwidth are the only known quantities) are taken into account \cite{powermonit}. Time-frequency methods have also been presented: the authors in \cite{bala} use the Wigner-Wille distribution to transform time series into two-dimensional maps in which ridges are looked for by using the Steger algorithm; a Student-like test on noise periodograms has been studied in order to detect non-stationary events \cite{mohanty}. More generally, we have developed a set of filters able to be used as on-line triggers \cite{arnaud,moriond,pra}, the most promising being the ``alternative linear fit filter'' (ALF) \cite{pra}, which aims at detecting a nonzero slope in data windows.

As burst signals are poorly modelled, only an accurate knowledge of the detector behaviour could allow to properly separate a real GW from non-stationarities in a single antenna; therefore, one will be more confident about the reality of a candidate event in case of a significant coincidence between several interferometers, or between an interferometer and other types of detector (neutrinos, $\gamma$-rays satellites...). Many studies in the literature assume such multi-detections as the starting point of the analysis and deal with the ``inverse problem'', i.e. how to extract informations -- source location in the sky, GW waveforms, astrophysical parameters -- from these triggered data \cite{gursel,jk5,pai1,finn}. In this article we will not address this question which will be the main topic of forthcoming papers aiming at developing coincidence strategies for the detection of bursts with a network of both GW antennas and detectors sensitive to other radiations such as neutrinos.

Simulations of massive star collapses \cite{bona1,zwerger} give typical GW amplitudes too low to expect a likely detection of sources outside the Galaxy; more optimistic estimates have been computed recently for the merging phase of two neutron stars \cite{ruf} or a neutron star and a black hole \cite{janka}, whose coalescences could be detected until 10 Mpc given the planned sensitivity of the first generation of detectors. In both cases, as the GW amplitude is just above the noise level, the signal to noise ratios (SNR) are nevertheless expected to be low, all the more that the wave amplitude detected in a given antenna depends on its location and orientation with respect to the source \cite{thorne87}: when the source direction is not orthogonal to the detector plane, the response to the GW is not optimal and so the SNR is decreased.

Therefore the feasibility and the efficiency of coincidences between interferometers currently under construction worldwide is an important question: are these coincidences likely or not? This problem has attracted so far little attention in the literature whereas it should be the starting point of the study of coincidence strategies: a precise knowledge on how a network of detectors surveys the sky is essential in defining the best possible methods to analyse their data.

This paper addresses this topic from a statistical point of view by performing Monte-Carlo simulations of the detection process in a network of antennas. After having described the model used in the computations, we briefly recall the structure of the beam pattern functions - describing the interaction between a GW and a detector - and extract from them some information on the accuracy with which a particular sky direction is surveyed by a given interferometer. We will use the celestial sphere coordinates, i.e.  the right ascension $\alpha$ and the declination $\delta$ to label sources. Sky maps will be constructed to characterise the detection power of each antenna and any combination of them.

Next, we introduce a quantity independent of any detector, $\rho_{\text{max}}$, characterising a GW strength; the simulation results -- detection probabilities for different configurations in the network: single antenna, twofold coincidence, threefold coincidence -- are then presented either as function of $\rho_{\text{max}}$ and $(\alpha,\delta)$ on a sky map, or simply versus $\rho_{\text{max}}$ after averaging over all the sky directions.

Finally, the timing accuracy is strongly connected to the detection problem: reconstructing the source location in the sky or performing coincidences with other types of detectors ($\gamma$-rays, neutrinos...) requires a good knowledge of the GW event timing in the interferometer(s). Therefore, we also present some results about the timing precision one can expect for typical burst detection before concluding.

\section{Modelling the detection process in the network of interferometers}

In this section we present the different hypothesis used in the Monte-Carlo simulations whose results are presented below: first about the network of detectors; then we give some assumptions about the GW burst and the calibration of its amplitude; finally, we detail the detection process itself.

\subsection{The network}

Each detector $D_i$ is described by a set of four angles: its latitude $l_i$, its longitude $L_i$, the angle between the two arms $\chi_i$ and an orientation angle $\gamma_i$ defined as the angle between the local South direction and the bisecting line of the detector arms counted counterclockwise. The values of these angles can be found in Table~I extracted from \cite{jk1}. 

\begin{center}
\begin{tabular}{|c|c|c|c|c|}
\hline Detector & Latitude $l$ & Longitude $L$ & Arms 'separation' $\chi$ & 'Azimuth' $\gamma$ \\
\hline VIRGO & 43.6 & -10.5 & 90.0 & 206.5 \\
\hline LIGO Hanford & 46.5 & 119.4 & 90.0 & 261.8 \\
\hline LIGO Livingston & 30.6 & 90.8 & 90.0 & 333.0 \\ 
\hline GEO600 & 52.3 & -9.8 & 94.3 & 158.8 \\
\hline TAMA300 & 35.7 & -139.5 & 90.0 & 315.0 \\
\hline
\end{tabular}
\end{center}
\centerline{Table I: detector data, all angles given in degrees}

The interferometer noise components are assumed to be white (after whitening \cite{cuo,cuo2} and applying line removal \cite{sintes} methods), Gaussian, broadband, additive and uncorrelated; lacking realistic sensitivities validated by measurements, -- these informations will appear when detector runs start -- we also assume they all have same noise standard deviation $\sigma$\cite{note1} and sampling frequency $f_0$.

\subsection{The gravitational wave burst and the source direction}

As a GW signal, we use an one-parameter set of Gaussian bursts
\begin{equation}
G_\tau(t)=K\exp\left(-\frac{t^2}{2\tau^2}\right)
\end{equation}
where the width $\tau$ is taken between 0.1 ms and 10 ms and $K$ is a scaling factor. Therefore, the maximum SNR (corresponding to an optimal detector orientation with respect to the source direction \cite{note2} and by using Wiener filtering), referred as $\rho_{\text{max}}$ in the following, is given by \cite{arnaud}:
\begin{equation}
\rho_{\text{max}}=K\frac{\pi^{1/4}\sqrt{f_0\tau}}{\sigma}
\label{eq:K}
\end{equation}
By assuming a value for $\rho_{\text{max}}$, it is straightforward to compute $K$ from Eq. (\ref{eq:K}) and so to characterise the amplitude of the GW burst independently of any detector.

In addition to the two coordinates $\alpha$ and $\delta$ previously introduced, a third variable is necessary to determine completely the GW: the polarisation angle $\psi$, one of the Euler angles describing the wave coordinate system in the TT gauge with respect to the celestial frame. In the simulations, unless specified otherwise, the GW sources are assumed to be uniformly distributed over the sky ($\alpha\in[-\pi;\pi]$, $\sin\delta\in[-1;1]$, $\psi\in[-\pi;\pi]$).

Finally, one has to notice that the arrival time of the GW in the detectors is a priori not synchronised with the sampling which will cause some losses in SNR and in timing accuracy for very narrow bursts.

\subsection{The detection procedure}

The detection method used is Wiener filtering -- correlation with the known signal itself -- for sake of simplicity: it is an approximation of no consequence as an earlier paper \cite{arnaud} showed that the one-dimensional parameter space $\left[\tau_{\text{min}}=\text{ 0.1 ms};\tau_{\text{max}}=\text{ 10 ms}\right]$ can be covered by a discrete lattice of Gaussian filters ensuring a detection with a mismatch SNR loss lower than 1 \%. The threshold value on the filter output is set to $\eta=4.89$ which corresponds to a false alarm rate of $10^{-6}$, i.e. 72 per hour for the VIRGO value of the sampling frequency $f_0$. One claims a detection if the filter output overcomes the threshold at least once in the analysis window. Decreasing the false alarm rate ($10^{-7}$, $10^{-8}$...) will not dramatically increase the threshold, so the results presented in the following remain essentially the same.

\section{Averaged beam pattern sky map}

\subsection{Beam pattern functions}

In the frame associated to the GW (wave direction being $z$ by convention), in the TT gauge, the spatial metric perturbation is given by
\begin{equation}
H(t)=
\begin{pmatrix}
h_+(t) & h_\times(t) & 0 \\
h_\times(t) & -h_+(t) & 0 \\
0 & 0 & 0
\end{pmatrix}
\end{equation}
with $h_+$ and $h_\times$ corresponding to the two independent wave polarisations. By assuming the detector size small compared to the reduced wavelength of the GW \cite{thorne87} and by following the treatment presented e.g. in \cite{jk5,jk1} it is straightforward to show that the response $h(t)$ of an interferometric detector to this wave is a linear combination of the two polarisations: 
\begin{equation}
h(t)=F_+(t)h_+(t)+F_\times(t)h_\times(t)
\end{equation}
The corresponding weighting factors are called beam pattern functions; they have values in the range [-1;1] depending on the longitude and the latitude of the detector location, as well as its orientation, the angle between the interferometer arms $\chi$, the sky coordinates $(\alpha,\delta)$ of the source and the wave polarisation angle $\psi$. 
Still following the notations and the analysis of \cite{jk1} one can compute the general expression of the beam pattern functions:
\begin{equation}
\begin{pmatrix}
F_+(t) \\ F_\times(t)
\end{pmatrix}
=
\sin\chi
\begin{pmatrix}
\cos2\psi & \sin2\psi \\
-\sin2\psi &\cos2\psi
\end{pmatrix}
\begin{pmatrix}
a(t) \\ b(t)
\end{pmatrix}
\end{equation}
The factor $\sin\chi$ recalls that the best response is achieved for detectors with orthogonal arms (like for instance VIRGO and the two LIGO interferometers); the $a$ and $b$ factors depend neither on $\psi$ nor on $\chi$ and so $2\psi$ appears like a rotation angle and $\sin\chi$ like a scale factor for the beam pattern functions.

Due to the Earth proper rotational motion the sky is in apparent motion with a period equal to a mean sidereal day. Therefore, the beam pattern functions associated to a source location ($\alpha$,$\delta$) depend also on the UT time $t$; let us introduce the local sidereal time $T(t)$ for a detector of longitude $L$:
\begin{eqnarray}
T(t) &=& \kappa t + T_{\text{Greenwich}}(0) - L\\
\text{with } \kappa &\approx& 1.0027379 \times 15^\circ/\text{hour} \nonumber
\end{eqnarray}
$T_{\text{Greenwich}}(0)$ is the Greenwich sidereal time at 0h UT and the minus sign before $L$ comes from the fact that longitudes are counted positive westwards.
Then one can define the local hour angle of the source:
\begin{equation}
{\frak H}(t) = T(t)-\alpha = \kappa t - ( \alpha + L ) + T_{\text{Greenwich}}(0)
\end{equation} 

Extensive calculations yield to compute the complete expressions of $a$ and $b$ (depending on $\alpha$, $\delta$, $l$, $L$, $\gamma$ and $t$); we recall them for sake of completeness even if they are completely equivalent to those presented in section 2.1 of \cite{jk1} (see also references therein):
\begin{eqnarray}
a(t)=&-&\frac{1}{16}\sin2\gamma\,(3-\cos2l)(3-\cos2\delta)\cos2{\frak{H}}(t)-\frac{1}{4}\cos2\gamma\,\sin l\,(3-\cos2\delta)\sin2{\frak{H}}(t) \nonumber\\
&-&\frac{1}{4}\sin2\gamma\,\sin2l\,\sin2\delta\,\cos{\frak{H}}(t)-\frac{1}{2}\cos2\gamma\,\cos l\,\sin2\delta\sin{\frak{H}}(t)-\frac{3}{4}\cos2\gamma\,\cos^2l\,\cos^2\delta \\
b(t)=&-&\cos2\gamma\,\sin l\,\sin\delta\,\cos2{\frak{H}}(t)+\frac{1}{4}\sin2\gamma\,(3-\cos2l)\sin\delta\,\sin2{\frak{H}}(t) \nonumber \\
&-&\cos2\gamma\,\cos l\,\cos\delta\,\cos{\frak{H}}(t)+\frac{1}{2}\sin2\gamma\,\sin 2l\,\cos\delta\,\sin{\frak{H}}(t)  
\end{eqnarray}

\subsection{Detectors sky maps}

The main interest of studying beam pattern functions is to characterise the sensitivity of an interferometric detector, given for different directions in the sky, independently of its noise curve. This can be done by performing a quadratic average of the beam pattern amplitude over the unknown polarisation angle $\psi$. It leads to a $\psi$-independent quantity $\overline{F}(t)$ \cite{note3} given by
\begin{equation}
\overline{F}(t) = \frac{\sin\chi}{\sqrt{2}}\sqrt{a^2(t)+b^2(t)}\in\left[0;\frac{1}{\sqrt{2}}\right]
\end{equation}

We choose to present the variations of $\bar{F}$ in a two-dimensional contour plot figure -- the ``sky map'' -- where any direction is located by a couple $(\alpha,\sin\delta)\in[-\pi;\pi]\times[-1;1]$. For a given detector, $\overline{F}$ depends not only on $\alpha$ and $\delta$ but also on the time $t$. For sake of simplicity, each sky map shown below in the paper is determined at the time $t=t_0$ defined by $\kappa t_0 + T_{\text{Greenwich}}(0)=0\,[2\pi]$; to have results at another time $t$ one can simply imagine the map on a cylinder whose axis is along the $\sin\delta$ direction and to rotate it by $\delta{\frak H}(t)={\frak H}(t)-{\frak H}\left(t_0\right)$.

\begin{figure}
\centerline{\epsfig{file=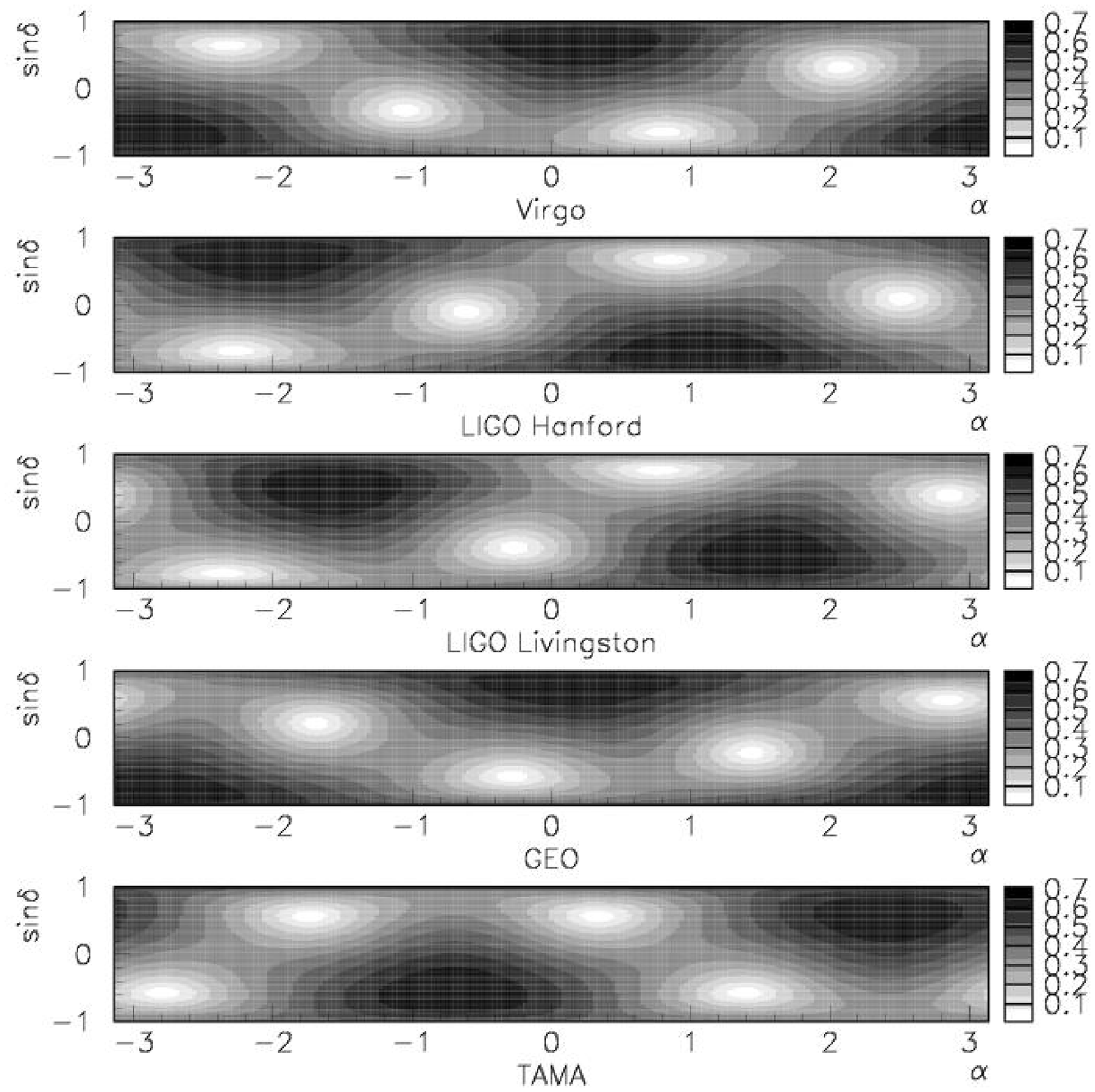,width=20cm}} 
\caption{Comparison of the averaged beam pattern maps averaged over the polarisation $\psi$ for the three detectors: VIRGO, LIGO Hanford and LIGO Livingston.}
\label{figure1}
\end{figure}

Figure \ref{figure1} compares the sky maps of the $\psi$-averaged beam pattern function $\overline{F}$ for the five interferometers currently in development in the world: VIRGO, the two LIGO antennas, GEO600 and TAMA300. In each case, the $\overline{F}$ averaged value over the whole sky is 0.42 and the RMS is 0.16; each map presents 2 maxima corresponding to directions which are orthogonal to the detector plane and 4 minima (direction of the bisector of the two arms and its three images by rotations of angles $\pi/2$, $\pi$ and $3\pi/2$ in the detector plane) included in a large ``valley'' where $\overline{F}$ remains quite small.

To simplify the analysis presented in this paper we now consider only three of these antennas as this is the minimum number required in order to be able to reconstruct the source direction in the sky in case of a full coincidence. We keep the two LIGO interferometers as they have been planned to work coherently by construction and as they have the same design sensitivities; VIRGO and GEO600 have almost identical $\overline{F}$ sky map (at least for the areas where the averaged beam pattern function is maximal) and so keeping one of the two (retaining VIRGO which is expected to have a sensitivity similar to LIGO) is enough to compute the main properties of an American-European network. Finally, despite that TAMA300 sky map is different from the other ones, its expected sensitivity is too small -- with respect to the other antennas previously mentioned -- to help the network detection. So, in the following, the network chosen is made of three detectors: VIRGO (V) and the two LIGO interferometers in Hanford (H) and Livingston (L); the selected antennas are those expected to achieve the best broadband sensitivities. Nevertheless the results presented in this paper can be easily extended to accommodate additional interferometers into the network.

The regions of maximum $\overline{F}$ are rather close for the two LIGO interferometers but quite different from those of VIRGO. Therefore one can expect on one hand a good complementarity between the three interferometers for a single detection -- a clear advantage to maximise sky coverage. On the other hand, the ability to perform coincidences between the 3 detectors is expected to be significantly reduced -- at least for small SNR values.

\subsection{Daily averaging}

As mentioned before, the two-dimensional maps previously defined are also function of time. Therefore, another way to present the information they contain is to average $\overline{F}$ over one sidereal day (and so over $\alpha$) to keep only the dependence in the declination variable $\sin\delta$. 

\begin{equation}
\left(\overline{F}\right)^{\text{averaged}}(\sin\delta)=\frac{1}{2\pi}\int_{-\pi}^{\pi}\text{d}\alpha\, \overline{F}(\alpha,\sin\delta)
\end{equation}

Figure \ref{figure2} presents the graphs of $\left(\overline{F}\right)^{\text{averaged}}$ for the three detectors VIRGO, LIGO Hanford and LIGO Livingston. One can note that the vertical axis scale is zero-suppressed to enhance the variations which are in fact small ($\lesssim30\%$) around the mean value of 0.42. Due to the successive averages, these plots mainly depend on the detector's latitude which explains why VIRGO and LIGO Hanford curves are quite similar. To make this figure more concrete, vertical lines are given corresponding to the location of some centres of galaxies where GW sources are expected to be found: the Galaxy, the Magellanic clouds, M31 Andromeda and M87 in the Virgo cluster. For the galactic centre and M31, the three detectors are roughly identical whereas Livingston is better for M87 and VIRGO/Hanford for the Magellanic clouds.

\section{Detector configurations}

The detection process (exceeding or not a threshold) is not linear; therefore, it can only be studied by using Monte-Carlo simulations. So, in the following, no a priori average on $\psi$ is performed; we assume uniform distributions of sources in the sky which corresponds to the ranges of variables quoted in Table II. The results are twofold:
\begin{itemize}
\item Detection efficiencies versus $\rho_{\text{max}}$ averaged over $\alpha$, $\sin\delta$ and $\psi$;
\item Detection sky maps, i.e. detection probability versus $(\alpha,\sin\delta)$ for a given $\rho_{\text{max}}$, averaged on $\psi$.
\end{itemize}
Various configurations are studied, from a single interferometer to the search for full coincidences between VIRGO and the two LIGO antennas.

\subsection{Single detector}

\subsubsection{The example of VIRGO}

As shown in section III-A, the amplitude of the GW interacting with a detector has to be multiplied by the beam pattern function with respect to the ideal case (optimal incidence and polarisation). Therefore the signal to noise ratio is lowered by the same factor: let us call $\rho_{\text{real}}$ the product $F \times \rho_{\text{max}}$ where $F$ is the beam pattern function. Figure \ref{figure3} presents the detection efficiency as function of the real signal to noise ratio $\rho_{\text{real}}$ for a linear filter with a $10^{-6}$ false alarm rate \cite{note5}. The 50\% efficiency is reached for $\rho_{\text{real}}=\text{threshold}\approx5$ and the detection is almost always successful for real SNR higher than 7 (efficiency higher than 98\%). Note that this curve obviously does not depend on the beam pattern.

Adding the effect of the non optimal detector response gives graphs such as those shown on Figure \ref{figure4}. For two values of $\rho_{\text{max}}$ - 10 and 20 - and a Gaussian width of $\tau=1$ ms, one sees 
\begin{itemize}
\item continuous line: the distribution of $\rho_{\text{real}}$ for $\rho_{\text{max}}$ constant; this curve shape is only due to the beam pattern;
\item dashed line: the fraction of events really detected, computed by including detector efficiency from Figure \ref{figure3}.
\end{itemize}
From 0 to $\frac{\rho_{\text{max}}}{2}$ the distribution of $\rho_{\text{real}}$ is flat; for higher values, it decreases monotonously up to the maximal signal to noise ratio. Adding the noisy detection process considerably lowers the number of events detected at small real SNR but keeps almost all those with  $\rho_{\text{real}}\ge7$. By comparing the number of detected events with the total number of events generated by the Monte-Carlo simulation, one can compute the detection efficiencies: 32\% for $\rho_{\text{max}}=10$, 66\% for $\rho_{\text{max}}=20$.

Simulating the real detection process, one obtains Figure \ref{figure5} which shows the detection probability versus $\rho_{\text{max}}$ for different values of $\tau$. As expected, this probability only depends on the signal to noise ratio which is the relevant variable in signal analysis. Due to the beam pattern functions, the detection efficiency remains low even for large optimal signal to noise ratios: with $\rho_{\text{max}}$=10 the detection efficiency is only about 30\%, it reaches 50\% for $\rho_{\text{max}}$=14 and 90\% for -- unlikely -- high values.

From those graphs one can conclude that including beam pattern functions in data analysis has major consequences on detection efficiency by reducing greatly the detection probability even for high $\rho_{\text{max}}$.

The previous results deal with averaged efficiencies around the whole sky; of course, the detection probability depends on the relative position of the source with respect to the interferometer. Figure \ref{figure6} presents the detection efficiencies versus $(\alpha,\sin\delta)$ for three values of $\rho_{\text{max}}$. One can note that these three maps are highly correlated with the upper graph of Figure \ref{figure1}: $\overline{F}$ is thus a good -- and easily computable -- estimator of the detection probability in a given sky direction, especially in the low SNR region where are likely to be located the first detected events.

In the first case $\left(\rho_{\text{max}}=10\right)$ detections are possible for incoming wave directions almost orthogonal to the detector plane reducing considerably the detection rate: a large fraction of the sky remains invisible. For $\rho_{\text{max}}=15$ and $\rho_{\text{max}}=20$, the detection probability is more uniform over the sky; the four minima are still there but they cover now narrower regions whose areas decrease as $\rho_{\text{max}}$ increases. In the last case, there are simply blind islands among regions where the detection efficiency is larger than 60\%.

Converting the results of these maps into fractions of sky with given detection probability leads to Figure \ref{figure14}. One can note that especially at large $\rho_{\text{max}}$ the fraction of sky covered decreases first slowly when the detection probability level increases and then more rapidly after some corner value. The reason for this evolution can be understood by looking at Figure \ref{figure6}, considering the large part of the sky where the efficiency is large and almost constant. The detection probability is higher than 30\% in 40\% of the sky for $\rho_{\text{max}}$=10; for $\rho_{\text{max}}$=15 it is more than 50\% in half of the sky and it is almost 60\% in 70\% of the sky for $\rho_{\text{max}}$=20. Therefore detections are likely for high maximal SNR $\rho_{\text{max}}$ in any direction but even for $\rho_{\text{max}}$=20, the efficiency never reaches 90\%.

\subsubsection{Daily averaged detection probability}

\begin{figure}
\centerline{\epsfig{file=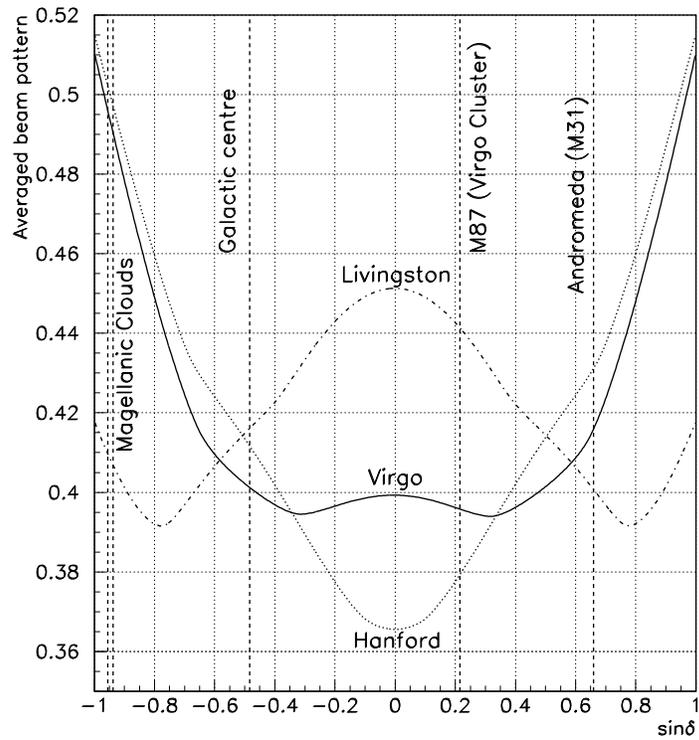,height=10cm}} 
\caption{Daily averaged $\overline{F}$ for the three detectors VIRGO and the 2 LIGO interferometers; the vertical dashed lines correspond to the location of some galaxy centres: the Galaxy, Magellanic clouds, Andromeda M31 and M87, one of the biggest components of the Virgo cluster.}
\label{figure2}
\end{figure}

\begin{figure}
\centerline{\epsfig{file=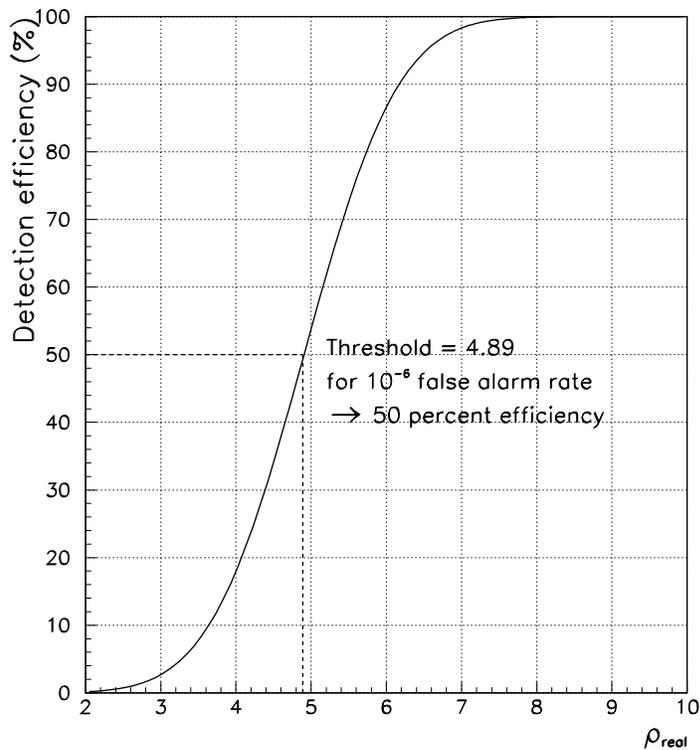, height=10cm}}
\caption{Detection efficiency versus $\rho_{\text{real}}$ for a false alarm rate of $10^{-6}$.}
\label{figure3}
\end{figure}

\begin{figure}
\centerline{\epsfig{file=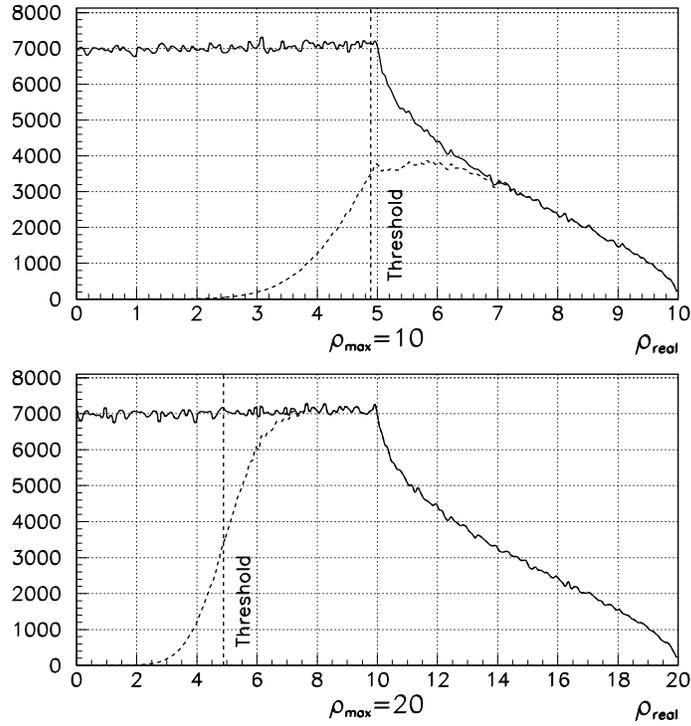, height=10cm}}
\caption{Distribution of the real SNR of the GW burst assuming $\rho_{\text{max}}$=10 or 20 and $\tau$=1 ms. The continuous line shows the SNR distribution only due to the beam pattern functions; the dashed line represents the final fraction of events detected by the filtering method computed by using the results shown in Figure \ref{figure3}.}
\label{figure4}
\end{figure}

\begin{figure}
\centerline{\epsfig{file=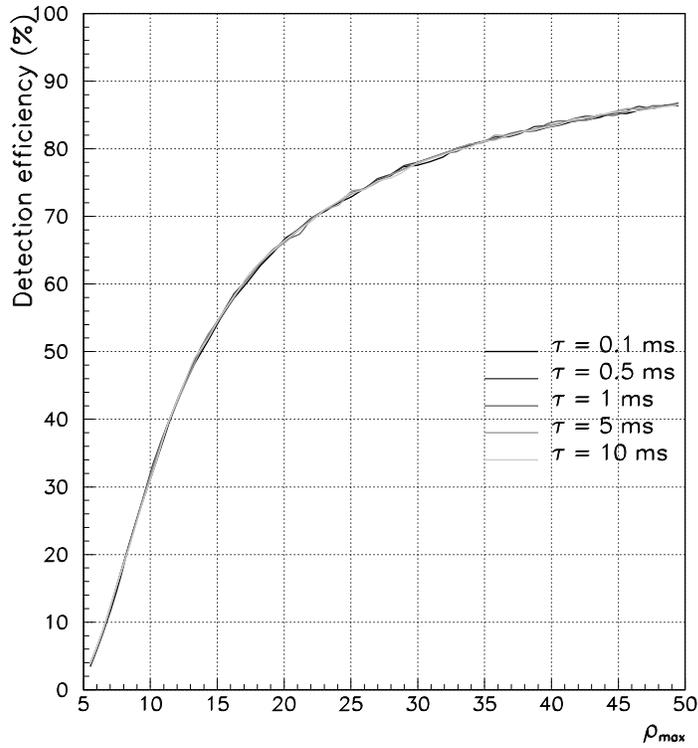, height=10cm}}
\caption{VIRGO detection efficiency (in \%) for different values of the width $\tau$ of the Gaussian signal as function of $\rho_{\text{max}}$.}
\label{figure5}
\end{figure}

\begin{figure}
\centerline{\epsfig{file=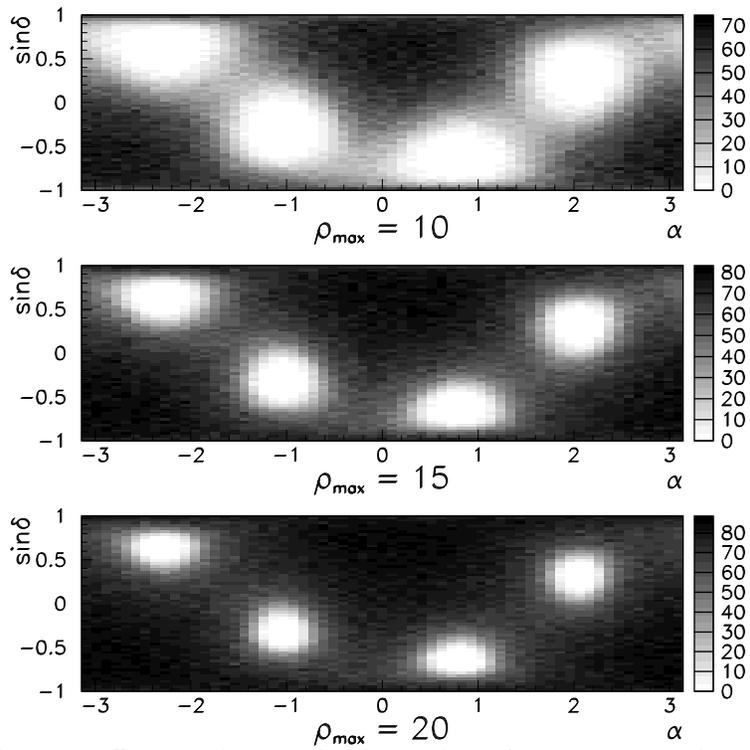, height=10cm}}
\caption{Comparison of detection efficiency sky maps for three values of $\rho_{\text{max}}$: 10, 15 and 20. The polarisation angle $\psi$ is randomly generated ensuring more realistic probabilities. Note the differences in the colour code on the various graphs.}
\label{figure6}
\end{figure}

\begin{figure}
\centerline{\epsfig{file=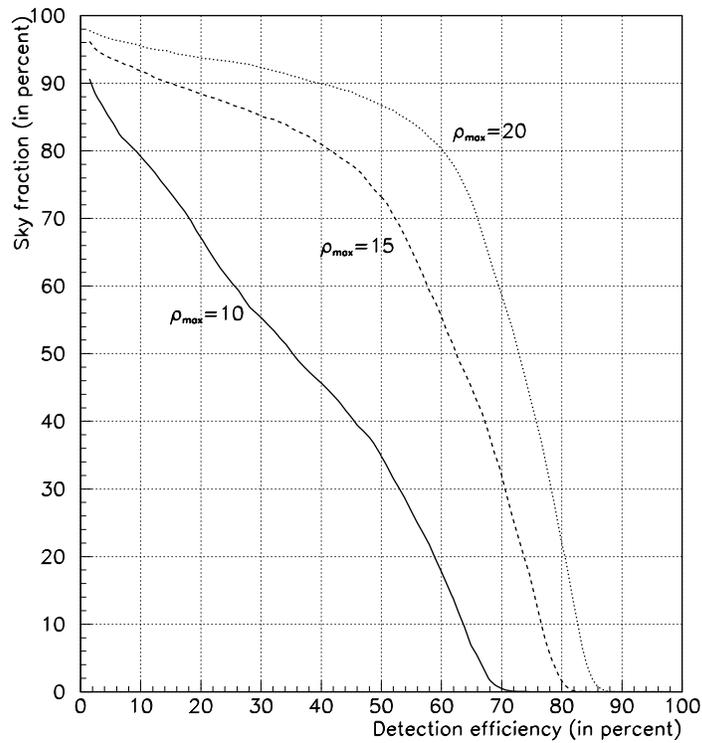, height=10cm}}
\caption{Fraction of sky (in \%) associated to a detection efficiency higher than a given value for $\rho_{\text{max}}=10$, 15 and 20.}
\label{figure14}
\end{figure}

\begin{figure}
\centerline{\epsfig{file=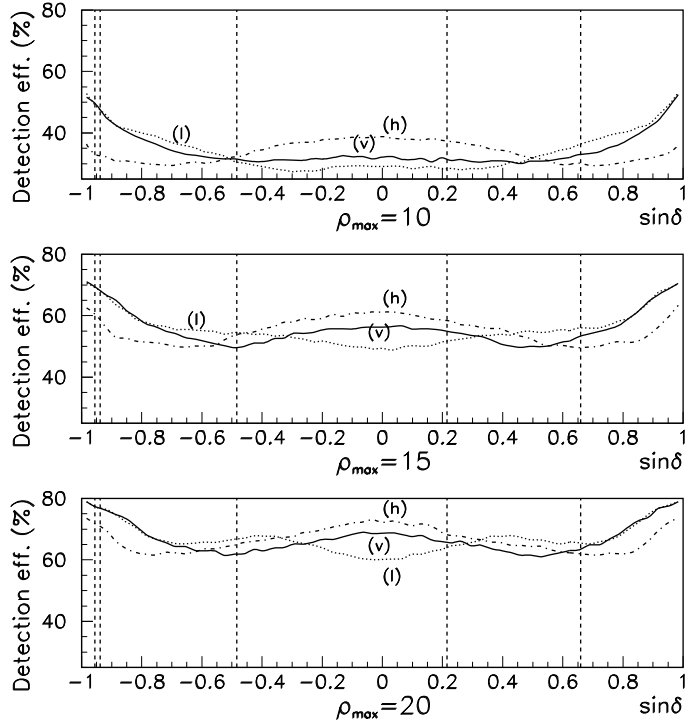, height=10cm}}
\caption{Detection efficiency (in \%) averaged over one day for the three detectors of the network and three values of $\rho_{\text{max}}$: 10, 15 and 20.}
\label{figure12}
\end{figure}

This analysis can also be done for the two LIGO interferometers; as expected, averaging the detection probability over the whole sky while taking $\rho_{\text{max}}$ constant gives exactly the same results. To compare the three interferometers, we simply average on $\alpha$ over one day. So, Figure \ref{figure12} shows the daily averaged detection efficiencies for each of the detectors and for three different values of the maximal signal to noise ratio, $\rho_{\text{max}}=10$, 15 and 20. As on Figure \ref{figure2}, the dashed vertical lines indicate some locations of galaxy centres. One can note that the detection probability is shifted to higher values for increasing $\rho_{\text{max}}$; nevertheless, the curve shapes do not change too much and remain close to those shown on Figure \ref{figure2} for the averaged beam pattern functions: LIGO Livingston has the best results for small values of $|\sin\delta|$ whereas LIGO Hanford and VIRGO are more efficient for large $|\sin\delta|$. But there remain some differences between the two Figures so that one should refine our previous statement: computing averaged beam pattern maps allow to have an idea of the detection probability in different sky areas but it is necessary to perform Monte-Carlo simulations including the detection process to compute the correct probabilities.

\subsection{Detector complementarity}

As we already checked that detection depends only on signal to noise ratio and not on the signal shape itself, we choose $\tau$=1 ms -- a typical burst duration for the Gaussian waveform we use -- in the following. In this section, we study the complementarity of the antennas, i.e. how their sky coverages complement each other, either for a single detection -- the ``OR'' strategy -- or for different types of coincidences -- the ``AND'' strategies. 

\begin{figure}
\centerline{\epsfig{file=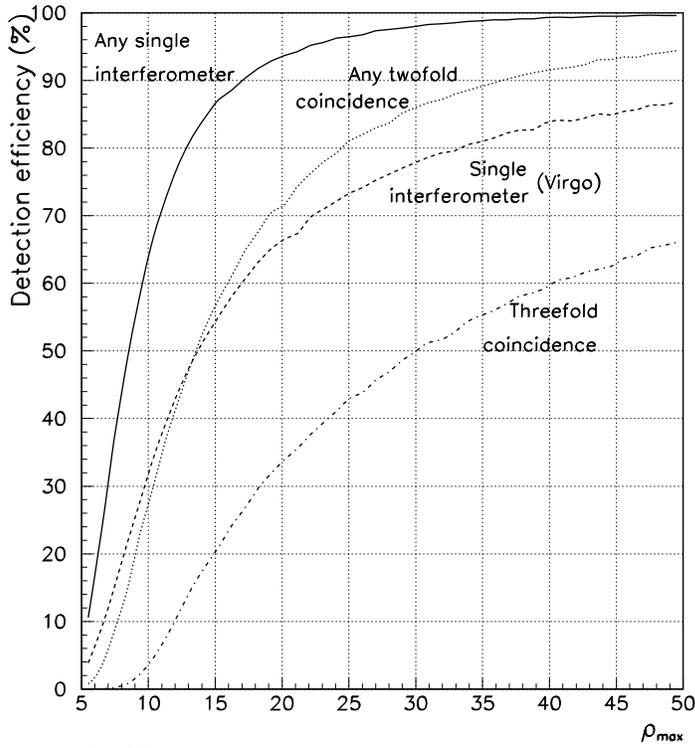, height=10cm}}
\caption{ Efficiency of detection (in \%) for various configurations of the VIRGO and the 2 LIGO interferometers network; continuous line: detection in at least one of the three interferometers; for comparison, dashed line recalls the detection efficiency in a single -given- one; dotted line: detection in at least two antennas; dotted dashed line, full coincidence in the three detectors.}
\label{figure7}
\end{figure}

Figure \ref{figure7} shows the detection probability versus $\rho_{\text{max}}$ for different configurations of detectors in the network; the curve already presented on Figure \ref{figure5} -- single detector efficiency -- is simply recalled for comparison. The continuous line shows the efficiency of detection in any single detector, i.e. the probability for a detection in at least one of the three interferometers. Network detection potential is clearly better than for a single antenna and the smaller the $\rho_{\text{max}}$ the higher the difference: more than twice more detections for $5\le \rho_{\text{max}} \le 10$, more than 1.5 below $\rho_{\text{max}}$=17. The 50\% efficiency is reached at $\rho_{\text{max}}$=8 and for $\rho_{\text{max}} \ge 15$ the probability is higher than 85\% on average, corresponding to a likely detection in most parts of the sky.

On the same Figure, one can also see the curve corresponding to a coincidence detection in at least two detectors among three (``OR'' of twofold coincidences). For small values of $\rho_{\text{max}}$, the twofold coincidence probability remains lower than the detection efficiency in a single interferometer but above $\rho_{\text{max}}=13$ -- where the two compared probabilities are about equal to 50\% -- it becomes more likely to trigger in two detectors. Finally, the probability of full coincidence in the three interferometers of the network is considered. It is quite small for $\rho_{\text{max}} \le 10$ and reaches useful levels only for very high maximal signal to noise ratios: the 50\% efficiency is reached only for $\rho_{\text{max}}=30$.

\begin{figure}
\centerline{\epsfig{file=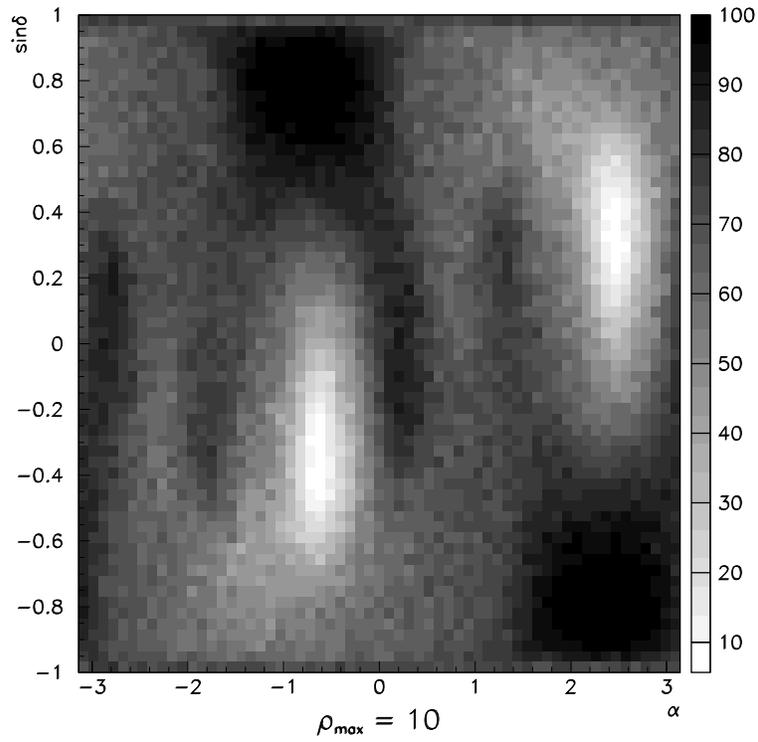, height=10cm}}
\caption{Sky map of the detection efficiency in at least one of the three interferometers; $\rho_{\text{max}}$ has been taken equal to 10.}
\label{figure8}
\end{figure}

To conclude this section, Figure \ref{figure8} presents the detection probability in at least one interferometer of the network (``OR'' strategy) as a function of the source location in the sky. The maximal signal to noise ratio is set to 10 and so this graph can be compared with the top map of Figure \ref{figure6}. The better efficiency achieved (about double, 67\% in average instead of 34\%) indeed corresponds to a more homogeneous detection probability over the sky. 

In this section and the next more details are given on the coincidence probabilities as simultaneous detection in different interferometers will ensure much higher confidence levels for candidate events.

\subsection{Twofold coincidences}

\begin{figure}
\centerline{\epsfig{file=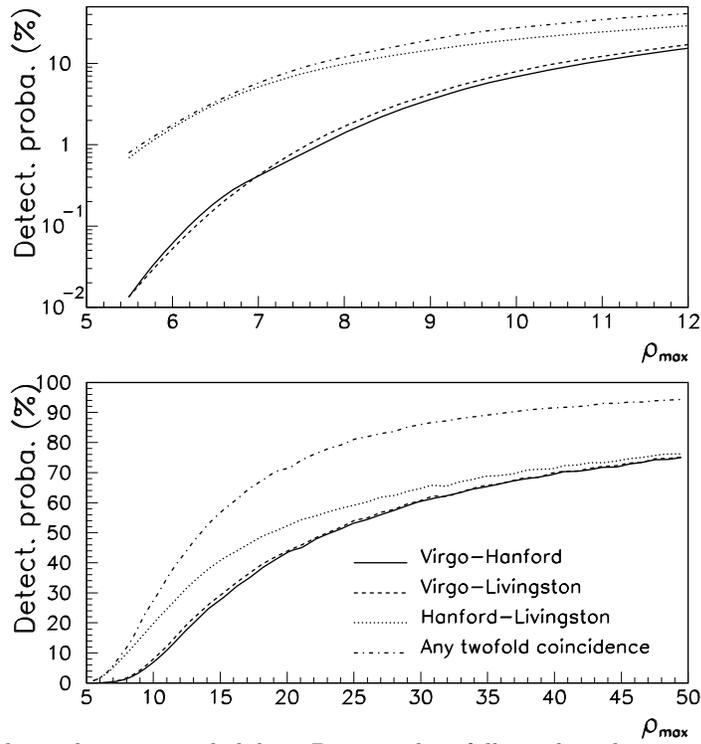, height=10cm}}
\caption{Twofold coincidence detection probability. Bottom plot: full graph with $\rho_{\text{max}}\in[5;50]$; top plot: zoom on small maximal signal to noise ratio values.}
\label{figure9}
\end{figure}

Figure \ref{figure9} shows the detection efficiencies for the three combinations of two detectors: VIRGO-Hanford, VIRGO-Livingston and Hanford-Livingston. By construction, the association between the two LIGO antennas is always more efficient than VIRGO and one of the LIGO interferometers. Nevertheless, the detection efficiency remains small: it is only 20\% for the two LIGO antennas with $\rho_{\text{max}}$=10. One can therefore conclude that simultaneous detection in two given detectors is unlikely for weak GW signals and becomes likely ($\approx40\%$) only for $\rho_{\text{max}}\ge15$. The last graph of this figure presents the twofold coincidence detection probability in the network, when triggering occurs in at least two interferometers among the three. For the smallest values of $\rho_{\text{max}}$ it corresponds to the two LIGO antennas case but then it increases more quickly: already at $\rho_{\text{max}}=10$ this probability is about 50\% larger than for the two LIGO interferometers.

\begin{figure}
\centerline{\epsfig{file=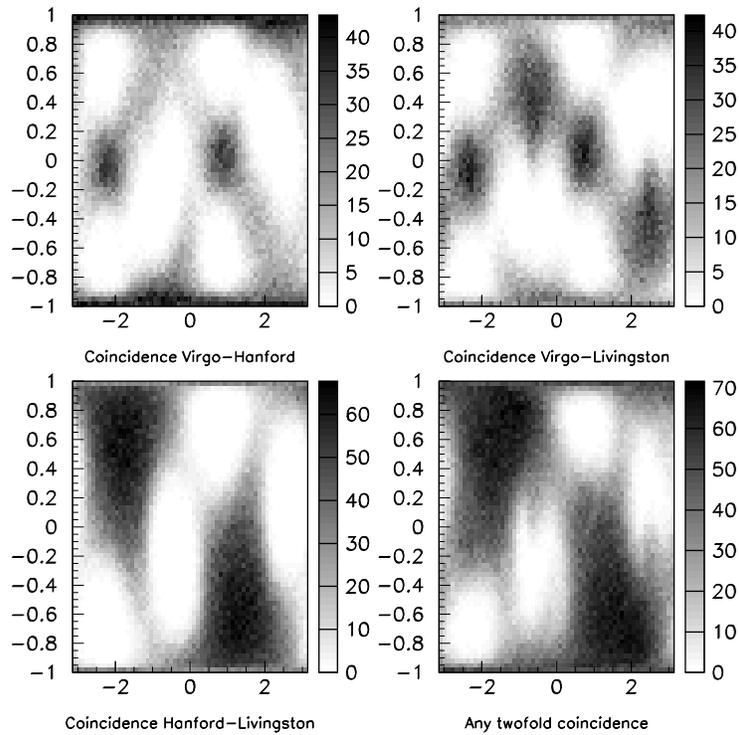, height=10cm}}
\caption{Twofold detection efficiency sky maps computed with $\rho_{\text{max}}$=10.}
\label{figure10}
\end{figure}

As the coincidence detection efficiency remains low for small values of $\rho_{\text{max}}$ it is instructive to investigate how it is distributed over the sky. Figure \ref{figure10} presents the detection efficiency of twofold coincidences versus $(\alpha,\sin\delta)$ for $\rho_{\text{max}}=10$ -- a likely value for the SNR of the first GW events detected. The three first maps are for the different twofold coincidences in the network while the fourth shows the detection probability in at least two detectors. One can see that despite the differences in the colour code the areas of high coincident detection probability are of small extent for the three configurations. In the last graph these regions are connected and the detection efficiency is higher, but nevertheless a major part of the sky remains invisible for the two-detector coincidences for low -- but unfortunately realistic -- $\rho_{\text{max}}$ values.

\subsection{Threefold coincidences}

From the conclusions of the previous sections it is clear that a simultaneous detection in the three detectors is not likely unless the optimal signal to noise ratio is large. Therefore it appears difficult to reconstruct GW astrophysical informations from a source at the expected sensitivity level provided by the first generation of detectors.

\begin{figure}
\centerline{\epsfig{file=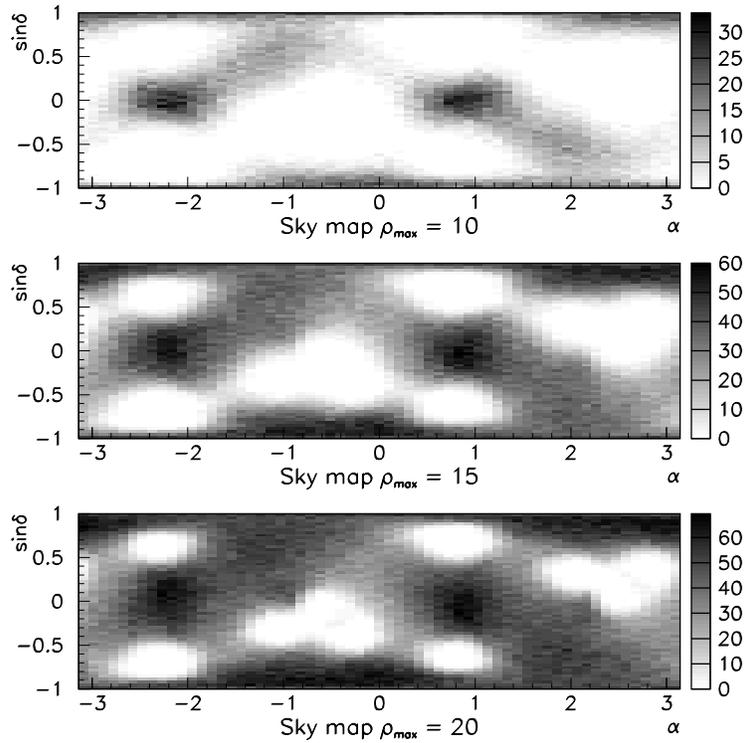, height=10cm}}
\caption{Threefold coincidences: sky maps for three values of the optimal signal to noise ratio: $\rho_{\text{max}}=10$, 15, 20; note the differences in the colour code.}
\label{figure11}
\end{figure}

Nevertheless, Figure \ref{figure11} shows the detection efficiency sky maps for values of the maximal signal to noise ratio of 10, 15 and 20 respectively. In the first case ($\rho_{\text{max}}=10$), non zero efficiency is concentrated in two small regions corresponding to the common visible areas of the LIGO-VIRGO, but even in these parts of the sky the detection efficiency is lower than 30\%. In the two other cases, the distribution is more uniform with unfortunately some large areas which remain invisible -- though they decrease as $\rho_{\text{max}}$ increases.

\section{Timing reconstruction}
Determination of the absolute timing of a detected GW burst is an important question. Firstly, coincidences in different antennas require time correlations between the respective detected signals. Secondly, the reconstruction of the source location is only based on arrival times detected in three detectors located in different places on Earth and the better the precision, the smaller the angular error box in the sky. Thirdly, coincidences with other types of detectors will also be based on time correlations. An application of the later using neutrinos can be found in \cite{MD}.

Let $\Delta t$ be the time difference between the actual arrival time of the GW on the detector $t_{\text{real}}$ and the reconstructed time $t_{\text{detect}}$ corresponding to a detection with a given filter in the interferometer's output, when the SNR is maximal. 

\begin{equation}
\Delta t = t_{\text{real}} - t_{\text{detect}}
\end{equation}

Generally speaking, there could be an offset between $t_{\text{real}}$ and $t_{\text{detect}}$ specific to the filter algorithm but here we assume it to be zero as would be obtained with matched filtering whose output peaks when the signal and the filter overlap\cite{titithese}. Then the only significant parameter is the RMS of the distribution $\Delta t^{\text{RMS}}$. 

A priori $\Delta t$ can be split into two parts:
\begin{itemize}
\item $\Delta t_{\text{sampling}}$ due to the discrete data sampling of the experiments. For a signal of characteristic duration $\tau\gg\frac{1}{f_0}$ it is completely negligible as it is well approximated by an uniform distribution in the range $\left[-\frac{1}{2f_0};\frac{1}{2f_0}\right]$ whose standard deviation is $\frac{1}{\sqrt{12}f_0}\approx1.4\times10^{-2}$ ms e.g. for VIRGO. But for signals of small duration it can become significant by randomly dropping high amplitude parts of the GW whose consequence would be to trigger off maximum (or to loose the event).
\item $\Delta t_{\text{noise}}$: if the GW signal is embedded in detector noise, the precise location of the output highest value will depend on the actual noise time series; one expects this component to dominate for large $\tau$. From dimensional analysis it is clear that $\Delta t$ must be proportional to $\tau$ and so it is convenient to use the dimensionless quantity $\frac{\Delta t}{\tau}$.
\end{itemize}

\begin{figure}
\centerline{\epsfig{file=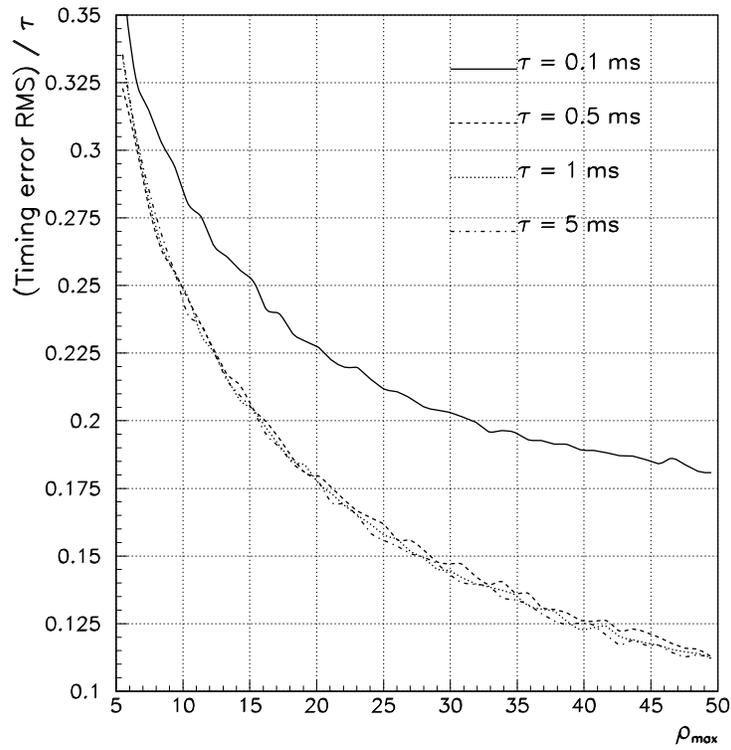, height=10cm}}
\caption{Normalised timing error RMS $\left(\frac{\Delta t^{\text{RMS}}}{\tau}\right)$, $\tau$ being the Gaussian width, versus the maximal signal to noise ratio $\rho_{\text{max}}$ and for five different values of $\tau$ between 0.1 and 10 ms.}
\label{figure13}
\end{figure}

Figure \ref{figure13} shows the normalised RMS $\frac{\Delta t^{\text{RMS}}}{\tau}$ as a function of $\rho_{\text{max}}$ for different values of the Gaussian width $\tau$. Apart from the case with $\tau=0.1$ ms which is sensitive to the finite sampling frequency as previously mentioned, all the other curves are identical. 

\begin{figure}
\centerline{\epsfig{file=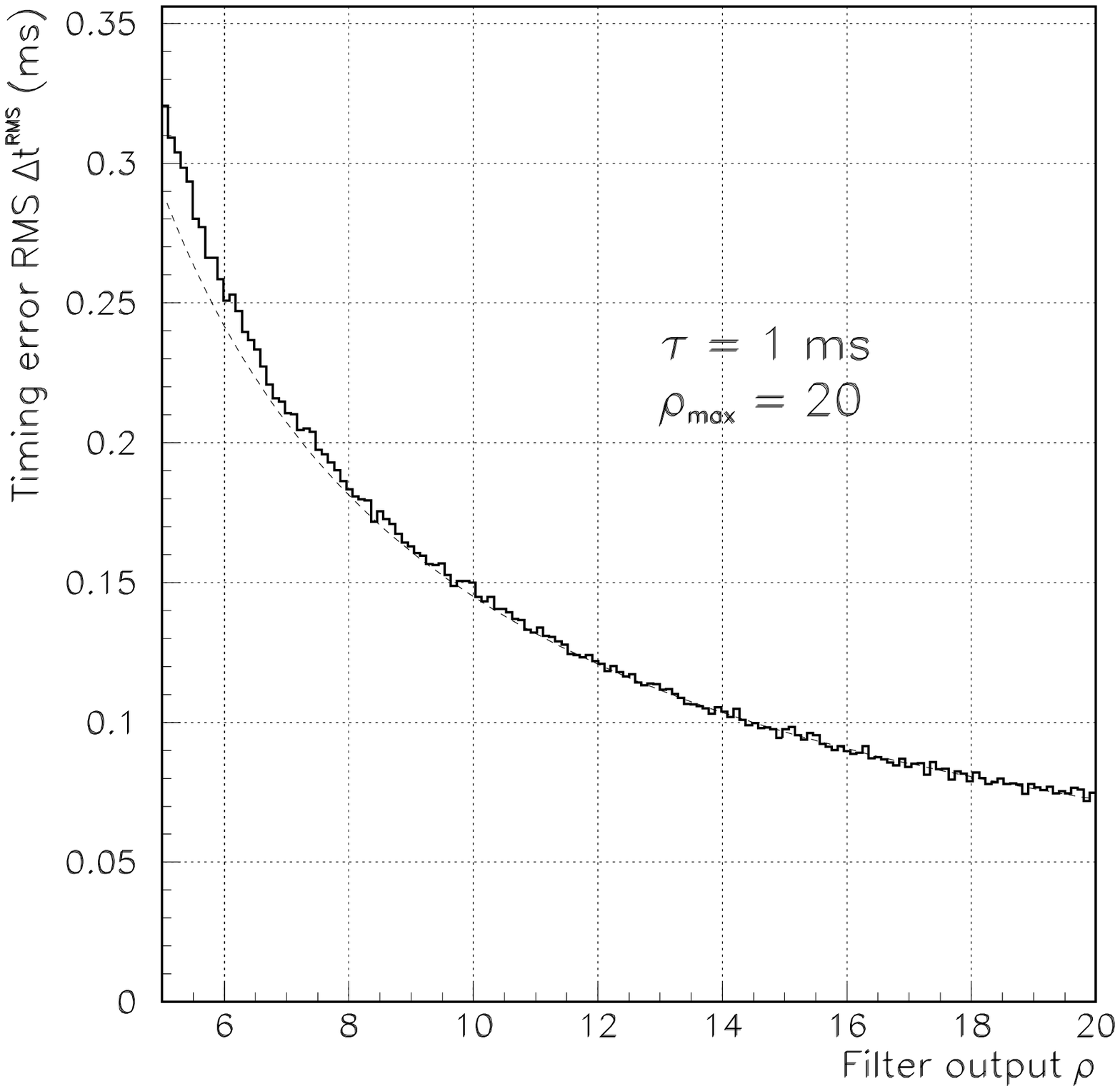,height=10cm}}
\caption{$\Delta t^{\text{RMS}}$ versus $\rho$ for $\tau=1$ ms and $\rho_{\text{max}}=20$; the dashed line shows the fit value.}
\label{figure15}
\end{figure}

From the data analysis point of view, it is mandatory to link the timing error with a measured quantity, the filter output $\rho$. The dependence of $\Delta t^{\text{RMS}}$ on this variable for $\tau$ values insensitive to the sampling rate can be represented by

\begin{equation}
\Delta t^{\text{RMS}}\approx \frac{1.45}{\rho}\left(\frac{\tau}{\text{1 ms}}\right)
\end{equation}   
The validity range of this equation is $\tau\gtrsim 0.2 \text{ ms}$ and the filter output $\rho \gtrsim 5-6$, as shown on Figure \ref{figure15} for a particular case, $\tau=1$ ms and $\rho_{\text{max}}=20$: the fit curve and the real one are in good agreement from $\rho=6$.

For $\tau=1$ ms and a filter maximal output of about 10, one has $\Delta t^{\text{RMS}}\approx 0.1$ ms which is well below the millisecond, the minimum level of precision suitable for coincidences with for instance neutrino detectors \cite{wip}.

\section{Conclusions}

This paper deals with the basic aspects of burst coincidence detection in a network of three gravitational wave antennas: VIRGO, LIGO Hanford and LIGO Livingston. It does not focus on the signal parameter reconstruction -- the inverse problem -- but rather on studying the coincidence detection probability to see whether this kind of event is likely or not. In this respect, the results obtained are somewhat disappointing: the detection probability in a given interferometer is strongly reduced by the beam pattern functions -- on the average only about 40\%. We also show that there is a good complementarity between VIRGO and the two LIGO interferometers for detection ( the 'OR' strategy efficiency is higher than 50\% for an optimal signal to noise ratio $\rho_{\text{max}}\gtrsim 8$), but coincidences are very unlikely for weak signals.

Concerning twofold coincidences, the two LIGO antennas configuration has a much better detection efficiency compared to VIRGO with one of the LIGO interferometers (a factor of two larger for $\rho_{\text{max}}=10$). Nevertheless, adding VIRGO and looking for any twofold coincidences allows the region of likely detection to be significally extended and the efficiency to be correspondingly increased: a 50\% enhancement is obtained for $\rho_{\text{max}}=10$, leading to a value averaged over the sky slightly higher than 30\%). Finally, threefold coincidences between detectors are quite unlikely below the value $\rho_{\text{max}}=30$ where the 50\% efficiency is reached.

As far as timing is concerned, the situation is quite satisfactory: Monte-Carlo simulations show that the RMS timing can remain below 1 ms even for low values of $\rho_{\text{max}}$, which is an interesting point for coincidence with other kind of detectors. Even if the detection itself remains unlikely, the timing will be accurate enough if the GW is seen by the detectors.

The basic aspects covered in this paper about beam pattern functions and timing accuracy apply equally well to all transient sources whose characteristic times are much shorter than a day; in particular, this is the case for the coalescing binary signals which are expected to last from one second to a few minutes in the detector bandwidths. So coincidence analysis of binary inspirals will face the same detection efficiency problem as discussed for the bursts. Apart from this inescapable fact the situation of the data analysis is different here since the known binary signals can be searched for through matched filtering. This opens the possibility of ``coherent'' rather than ``coincidence'' analyses, as proposed in \cite{pai1,pai2}, which are shown \cite{finn} to give better results than a simple coincidence analysis; thus, this method may compensate for part of the geometric effects presented in this paper by using in an optimal way all the information available in the detectors. This improvement is made at the cost of a large increase in computing power since the source sky coordinates have to be included in the template definition. Therefore, despite the fact that it is sub-optimal, coincidence analysis may be the only available tool in the first years of operation of the interferometers, even for binaries and especially for poorly modelled sources such as GW bursts.

One can a priori imagine two main strategies about their search:
\begin{itemize}
\item if the detector understanding is high enough to allow a proper elimination of non-stationary noise events, a burst detection is quite likely in at least one interferometer of the network since they cover the sky in a complementary way. Extending the number of sensitive antennas would further improve the detection efficiency. This situation is nevertheless not likely to arise in the first periods of data taking; it could even then be hopeless if the interferometer noise is not Gaussian. However, this strategy is well-adapted to coincidences with non-GW detectors if they have a negligible false alarm rate \cite{wip};
\item GW detection will in general require coincidences between interferometers. They may not be very frequent unless improvements are performed in the detector noise levels. Nevertheless specific strategies must be devised in order to maximise sky coverage and detection efficiency \cite{wip2}.
\end{itemize}

The results presented in this article give an overview of the GW burst detectability in a network of antennas focusing only on the detection efficiency. To go beyond, it is necessary to take into account the corresponding false alarm rates and to compare all the available associations of detectors sensitive to GW or other types of radiation such as neutrinos. Forthcoming papers will deal with these questions and also present new ideas about the ``inverse problem'' \cite{wip,wip2}.

\baselineskip = 0.5\baselineskip  




\begin{references}
\bibitem{note1} The exact value of $\sigma$ is of no importance as all the discussion of the paper is focused on SNR comparison. For example, by assuming a sampling frequency $f_0=20$ kHz (like in the VIRGO experiment) and a uniform one-sided spectral density $S_h\approx16\times10^{-46}/\text{Hz}$ (equal to the minimum of the foreseen VIRGO power spectrum), one finds from $\sigma\approx\sqrt{\frac{S_hf_0}{2}}$ an RMS noise value of $\sigma=4\times10^{-21}$.
\bibitem{note2} This choice corresponds to what Kip Thorne calls 'optimum source direction and polarisation' p. 367 of \cite{thorne87}. $\rho_{\text{max}}$ is computed from the Wiener filtering theory by using the formula $\rho_{\text{max}}^2=\frac{f_0}{\sigma^2}\int G_\tau(t)^2dt$.
\bibitem{note3} Due to this procedure it is no longer possible to separate ``$+$'' and ``$\times$'' polarisations and the averaged values are so equal for the two beam pattern functions. This definition of $\overline{F}$ may seem to be somewhat arbitrary but it is well correlated with the detection probability as shown by results given later in the paper.
\bibitem{note5} This efficiency is simply given by $\frac{1}{2}\erfc\left(\frac{\eta-\rho_{\text{real}}}{\sqrt{2}}\right)$ where $\eta$ is the corresponding threshold and $\erfc$ the error complementary function.
\end{references}
\end{document}